%Paper: hep-ph/9406408
%From: HEGYI@rmk530.rmki.kfki.hu
%Date: Tue, 28 Jun 1994 09:31 GMT+1

\magnification=\magstep1
\raggedbottom

 at  8truept
 at 10truept

\font\bo=cmbx10  scaled\magstep1
\font\tsc=cmcsc10

\hsize=15.truecm  \hoffset=1.truecm
\vsize=23.truecm  \voffset=1.truecm
\topskip=1.truecm \leftskip=0.truecm

\pageno=1
\footline={\hfill}
\headline={\ifnum\pageno=1\hfil\else\hss\tenrm-\ \folio\ -\hss\fi}

\let\cl\centerline  \let\rl\rightline   
\let\bs\bigskip  \let\ms\medskip  \let\ss\smallskip

%input psfig

\null
\rl{hep-ph/9406408}
\rl{June 1994}
\vskip2.truecm
{\bo
\cl{REMARKS ON}\ss
\cl{KOBA-NIELSEN-OLESEN SCALING}
}

\vskip1.truecm
\cl{\tsc S. Hegyi}

\footnote\ {\it hegyi@rmki.kfki.hu}

\baselineskip=11pt
{\it
\cl{KFKI Research Institute}
\cl{for Particle and Nuclear Physics}
\cl{of the Hungarian Academy of Sciences,}
\cl{H--1525 Budapest 114, P.O. Box 49. Hungary}
}

\vskip2.truecm

\baselineskip=12pt
\midinsert\narrower\narrower
\noindent
{\rm
\underbar{ABSTRACT}\hskip.4truecm
It is shown that there is a second properly normalized KNO scaling
function, $nP_n(n/\bar n)=\varphi(z)$,
which has certain advantages in the analysis of KNO scaling.
First, the $nP_n$ are not influenced by the statistical and
systematic uncertainties of $\bar n$ hence $\varphi(z)$ provides
more selective power than the original KNO scaling function
$\bar nP_n(n/\bar n)=\psi(z)$.
Second, the new scaling function
generates scale parameter $\sigma=1$ since it
depends only on the combination of $z$ and the
scale parameter of $\psi(z)$.
An analysis of $\varphi(z)$ is given using
$e^+e^-$ annihilation data
for charged particle multiplicity distributions.
}
\endinsert
\vskip2.truecm
%cl{Working paper}

\vfill\eject
\noindent
One of the most influential contributions to the analysis of
multiplicity distributions was made more than 20 years ago by
Koba, Nielsen and Olesen~[1]. They put forward the hypothesis that
at very high energies the probability distributions $P_n$ for detecting
$n$ final state particles exhibit a scaling law of the form
$$
	P_n={1\over\bar n}\,\psi\bigg({n\over\bar n}\bigg).\eqno(1)
$$
That is to say, the $\bar nP_n$ measured at different energies
(i.e. $\bar n$) scale
to the universal curve $\psi$ when plotted against the multiplicity
$n$ rescaled by the average multiplicity $\bar n$. This is the famous
Koba-Nielsen-Olesen (KNO) scaling hypothesis.
Defining the scaling variable $z\equiv n/\bar n$
the scaling function $\psi(z)$ must satisfy
two normalization conditions:
$$
	\int_0^\infty\psi(z)\,dz=1\eqno(2)
$$
and
$$
	\int_0^\infty z\,\psi(z)\,dz=1,\eqno(3)
$$
i.e. $\bar z=1$. Obviously, the moments ${\cal M}_q$ of $\psi(z)$
are independent of collision energy if Eq.~(1) is satisfied,
$$
\eqalign{
	{\cal M}_q&\equiv\int_0^\infty z^q\psi(z)\,dz\cr&=
	\overline{z^q}=\overline{n^q}\big/\bar n^q=
	\hbox{\rm const.}
}\eqno(4)
$$

Concerning the KNO scaling form of $P_n$ given by Eq.~(1) let us make
a further remark. For a mathematically specified
distribution $F(x)$ $\sigma$ is called a scale parameter if
$F(x)$ has the form~[2]
$$
	{1\over\sigma}\,f\bigg({x\over\sigma}\bigg),
	\qquad\sigma>0.\eqno(5)
$$
The parameter $\sigma$ characterizes the dispersion, that
how widely the values of $x$ are spread
around the typical value, reflecting thus the scale
of the distribution. From Eq.~(1) it is seen that in the KNO
scaling form of $P_n$ the average multiplicity serves as a scale
parameter. This coincides with the fact that if KNO scaling
holds the dispersion of $P_n$ is measured, up to a constant of
proportionality, by $\bar n$. The form Eq.~(5) is obeyed by the
scaling function too because the requirement $\bar z=1$ usually
generates a scale parameter $\sigma\not=1$ for $\psi(z)$.

In the phenomenological analysis of multiplicity distributions
one of the successfully applied probability laws is the
negative binomial~[3]. It has the form
$$
	P^{^{N\!B\!D}}_n=
	{{\cal P}(k,n)\over n!}\,
	\bigg({\bar n\over k}\bigg)^n
	\bigg(1+{\bar n\over k}\bigg)^{-n-k}
	\eqno(6)
$$
where $\bar n$ and $k$ are free parameters and
$$
	{\cal P}(k,n)=
	k\,(k+1)\ldots(k+n-1)=
	{\Gamma(k+n)\over\Gamma(k)}\eqno(7)
$$
denotes the Pochhammer polynomial for (possibly non-integer)
rising factorial powers~[4]. Eq.~(6) has a nice KNO scaling
form: in the limit $\bar n/k\gg1$ and $n/\bar n$ fixed
$$
	\bar nP^{^{N\!B\!D}}_n\sim\psi_{_G}(z)\eqno(8)
$$
where
$$
	\psi_{_G}(z)=
	{\cal N}\cdot\theta^kz^{k-1}\exp(-\theta z)\eqno(9)
$$
is the gamma distribution with normalization constant
${\cal N}=\Gamma^{-1}(k)$, scale parameter $\theta$ and
shape parameter $k$. $\psi_{_G}(z)$ may have a variety of shapes,
depending on the value of $k$, whereas $\theta$ reflects its scale,
the degree of scatter of $z$ around $\bar z$.
The moments ${\cal M}_q$ of the gamma distribution are given by
$$
	{\cal M}_q=\theta^{-q}\,{\cal P}(k,q).\eqno(10)
$$
{}From the requirement ${\cal M}_1=1$ expressed by
the normalization condition Eq.~(3)
we see that the scale parameter $\theta$ in Eq.~(9)
is enforced to satisfy
$$
	\theta={\cal M}_1\big|_{\theta=1}=
	{\cal P}(k,1),\eqno(11)
$$
i.e. $\theta=k$. The scale parameter coincides with the shape
parameter of the KNO scaling function. KNO scaling holds for energy
independent $k$.

Since the work of Koba, Nielsen and Olesen
testing the validity of the scaling hypothesis and
the analysis of the scaling function $\psi(z)$
are of permanent interest in multiparticle phenomenology~[5].
But there is a noteworthy fact that escaped attention:
besides $\psi(z)$ there is a second
properly normalized scaling function obeyed by the $P_n$.
The requirement Eq.~(3) for the first moment of
$\psi(z)$ defines a KNO scaling function normalized to $1$,
$\varphi(z)\equiv z\,\psi(z)$. This yields the scaling law
$$
	nP_n=\varphi\bigg({n\over\bar n}\bigg)\eqno(12)
$$
for the multiplicity distributions,
the $nP_n$ should lie on the universal curve $\varphi(z)$ if Eq.~(1)
holds. The moments of the two KNO scaling functions are related
by \hbox{${\cal M}^{(\varphi)}_q={\cal M}^{(\psi)}_{q+1}$}.

\ss

Investigating $nP_n$ instead of $\bar nP_n$
has the obvious advantage that
the statistical and systematic errors of $\bar n$
do not give contribution to the experimental uncertainty in
the shape of the scaling function.
Therefore $\varphi(z)$ can be more selective between various
theoretical predictions than $\psi(z)$. Another noticeable property
of $\varphi(z)$ arises in connection with Eq.~(5).
Since the KNO function $\psi(z)$
obeys this form, with $\sigma$ constrained by
Eq.~(3), the scaling law in Eq.~(12) can be written as
$$
	nP_n=\varphi\bigg({z\over\sigma}\bigg).\eqno(13)
$$
\goodbreak\noindent
The $nP_n$ depend only on the scaling variable $z$ rescaled by the
(energy independent) scale parameter $\sigma$ of $\psi(z)$.
Thus $\varphi(z/\sigma)$ provides a properly normalized KNO scaling
function with scale parameter $1$ and unconstrained ${\cal M}_1$.
The gamma distribution exhibits the scaling law Eq.~(13)
with $\sigma=\theta^{-1}$. Using the new scaling variable
$$
	w=z\theta={z\over{\ \overline{z^2}-1}^{\ }}\eqno(14)
$$
($\theta$ being restricted according to Eq.~(11))
one arrives at the scaling function
$$
	\varphi(w)={\cal N}\cdot w^k\exp(-w)\eqno(15)
$$
with ${\cal N}=\Gamma^{-1}(k)$. The $nP_n$ should fall onto the above
scaling curve if the $\bar nP_n$ follow a gamma distribution
with constant shape parameter $k$.

It is of interest to find the family of distributions obeying the
form Eq.~(5). On the basis of the results of
Ferguson [6] let us consider the following three different types of
probability laws on $[0,\infty)$
\ss
\itemitem{$a) $} Lognormal distribution
\
\itemitem{$b) $} Pareto distribution, possibly inverse power distribution

\itemitem{$c) $} Generalized gamma or Stacy distribution.
\ss\noindent
If $x$ is a random variable distributed according to $a)$,
$b)$ and $c)$ above, the random variable $y=\ln x$ will
have a
\ss
\itemitem{$a^\prime$)} Normal distribution

\itemitem{$b^\prime$)} Exponential distribution, possibly negative or
translated

\itemitem{$c^\prime$)} Generalized normal distribution
\ss\noindent
respectively. According to Ferguson these distributions
play an important role in statistics since
they are the only known ones that generate scale
parameter families Eq.~(5) of densities admitting complete sufficient
statistics when samples of size~$\geq2$ are available~[6].

Among the six probability laws $a)-c^\prime$) we turn our attention
to the generalized gamma or Stacy distribution.
The ordinary gamma, Eq.~(9), and thus the exponential
law $b^\prime)$ appear as a special case. Moreover $a)-b)$,
the lognormal and Pareto laws (the former one being very important
in multiparticle physics~[7])
can be obtained from~$c)$ in certain limits of
the parameters~[6]. We summarize some basic properties
of this distribution to be used later. Further
details can be found in the textbook of Johnson and Kotz~[8].
The generalized gamma distribution has the form
$$
	\psi_{_{G\!G}}(z)=
	{\cal N}\cdot\theta^kz^{\mu k-1}\exp(-\theta z^\mu)
	\eqno(16)
$$
with scale parameter $\theta$ and shape parameter $k$ as in
Eq.~(9) and normalization constant ${\cal N}=\mu\Gamma^{-1}(k)$.
The additional parameter, the exponent $\mu$, generalizes
the gamma distribution in a simple and useful manner: Eq.~(16)
corresponds to a power with exponent $1/\mu$ of a gamma variable.
This can be seen most clearly from the structure of the moments:
$$
	{\cal M}_q=\theta^{-q/\mu}\,{\cal P}(k,q/\mu),\eqno(17)
$$
the $q$th moment of the generalized gamma coincides with the
$(q/\mu)$th fractional moment of the ordinary gamma at fixed
$\theta$ and $k$. To get properly normalized KNO function
$\psi(z)$ having ${\cal M}_1=1$ the scale parameter
$\theta$ must satisfy
$$
	\theta={\cal M}_1\big|_{\theta=1}=
	{\cal P}^\mu(k,1/\mu).\eqno(18)
$$
If the $\bar nP_n$ follow a generalized gamma distribution
with energy independent shape parameter $k$ and exponent $\mu$
the $ nP_n$ obey the scaling law expressed by Eq.~(15) with the
corresponding normalization constant ${\cal N}$ and scaling variable
$$
	w=z^\mu\theta=
	{z^\mu\over{\ \mu\,\big(\,\overline{z^{\mu+1}}-
	\overline{z^{\mu}}^{\,}}\big)}\eqno(19)
$$
as can be checked through eqs.~(17) and (18). This gives back
Eq.~(14) for $\mu=1$. It is worth noticing that the exponent
$\mu$ is absorbed entirely into the rescaled variable $w$.

The above extension of the gamma distribution is not new
in the physical literature.
For a large class of nonlinear stochastic processes with
pure multiplicative noise Eq.~(16) arises as the stationary solution
of the corresponding Fokker-Planck equation~[9]. On the basis of
the previous work the generalized gamma distribution
was reviewed by Carruthers and Shih~[5]
pointing out that numerous theoretical
KNO scaling functions are special cases.
In its restricted form with ${\cal M}_1=1$ and $\theta$ given by
Eq.~(18) $\psi_{_{G\!G}}(z)$
was rediscovered by Krasznovszky and Wagner~[10]. In a series of
papers
they carried out a pioneering work in determining the magnitude of
the exponent $\mu$ for a large amount of data available in various
reactions, see~[11] and references therein.
Here we consider briefly the arguments made
by Dokshitzer about the higher-order perturbative QCD effects
on the shape of $\psi(z)$ in $e^+e^-$ annihilations [12].

Over the past years much attention has been focused on the description
of multiplicity distributions for hard processes in the framework of
quantum chromodynamics (for a recent review see ref.~[13]).
Although the calculations predicted KNO scaling, which is indeed the
case in $e^+e^-$ annihilations, the scaling function proved to fall off
exponentially in contradiction to the experimental results.
According to observations~[14] the tail of $\psi(z)$ decreases
faster than exponential.
In a recent attempt Dokshitzer was able to take into account more
precisely the energy conservation in parton cascades by involving
next-to-next-to-leading QCD effects in the determination
of $\psi(z)$ and its moments [12]. The
obtained KNO scaling function behaves according to
$$
	\psi(z)\propto\exp(-[Dz]^\mu)\eqno(20)
$$
for large $z$ with calculable constant $D$ and exponent
$$
	\mu={1\over1-\gamma\,\big(\alpha_s({\cal Q})\big)}.
	\eqno(21)
$$
In the above formula
$\alpha_s$ is the QCD running coupling constant corresponding
to the hardness scale ${\cal Q}$ and
$$
	\gamma\,\big(\alpha_s({\cal Q})\big)=
	{d\ln\bar n({\cal Q})\over d\ln{\cal Q}}\eqno(22)
$$
is the multiplicity anomalous dimension responsible for the energy
growth of the average multiplicity $\bar n$. In ref.~[12]
it was concluded that the better account of conservation laws
drastically reduces the higher-order correlations and thus the width
of $\psi(z)$. The exponent $\mu$ was found to be $\mu\approx1.6$
which agrees with the results obtained in~[11].

The comparison of eqs.~(20) and (16) suggests that the generalized
gamma distribution is well suited to characterize how
the higher-order corrections of
perturbative QCD affect the shape of $\psi(z)$. Let us take
Eq.~(9), the ordinary gamma with exponentially decreasing tail
as a reference distribution. One can model the suppression of
correlations by raising the gamma variable $z$ to the power
$1/\mu<1$. With increasing $z$ this procedure reduces more and more
the contribution of the exponentially falling tail
to the multiplicity moments $\overline{z^q}$.
The original distribution will be squeezed, decreasing
faster than exponential and having fractional multiplicity moments
with rescaled rank $q/\mu$ in Eq.~(10).
The same method can be applied in the opposite case too: one can
model possible enhancement of correlations by raising
the gamma variable $z$ to a power $>1$. Recall that
the exponent $\mu$ in Eq.~(16) has already been
determined for numerous reaction types~[11]. This can provide
valuable information on the degree to which the correlations are
enhanced/suppressed relative to the KNO scaling form, Eq.~(9),
of the negative binomial distribution.

Having derived a new KNO-type scaling law for the multiplicity
distributions, Eq.~(12), it is tempting to check its validity on
experimental data. We analysed the charged
particle multiplicity distributions in $e^+e^-$ annihilations from
the ALEPH, AMY, ARGUS, DELPHI, HRS, L3, OPAL and TASSO collaborations
in the c.m. energy range $\sqrt s=9.4-91$~GeV [14]. Our aim was
twofold. First, to determine how closely the $nP_n$ from
different energies follow a universal scaling curve $\varphi(z)$.
Second, to measure how our decisive power changes, in terms of
$\chi^2$ statistics, if $\varphi(z)$ is fitted to theoretical
predictions instead of $\psi(z)$.

Figure 1 displays the experimental data for $nP_n$ plotted against
$n/\bar n$ (since the points are densely populated no different
symbols are used to represent different data sets). According to
expectation, the $nP_n$ follow a unique scaling curve in agreement
with the observed KNO scaling behavior of the $\bar nP_n$~[14].
Another expected feature
in connection with $\varphi(z)$ is the increase of $\chi^2$ in fitting
procedures. The $nP_n$ are
not influenced by the statistical and systematic errors of
$\bar n$ hence the shape of $\varphi(z)$ is less uncertain
than that of $\psi(z)$.
Studying each data set separately we carried
out fits both to $nP_n$ and $\bar nP_n$
(by a theoretical $\varphi(z)$ and $\psi(z)$ respectively)
in order to measure the significance of this effect in
different circumstances.
We used Eq.~(16) for the theoretical $\psi(z)$
with normalization constant $2\,{\cal N}$
(because of the charged particle data),
scale parameter $\theta$ given by Eq.~(18) and exponent
$\mu$ fixed at the value $\mu=1.6$. According to
our results the increase in $\chi^2$ is substantial:
among the 11 analysed data sets it was found to be
$>50$~\% for 9 distributions. In the fitted values of the shape
parameter $k$ no significant change was observed.
The minimum and maximum increase in $\chi^2$ correspond to the
ALEPH and TASSO 34.8 GeV data respectively.
In the former case the variation is only marginal:
$$
	\chi^2_\psi\big/\hbox{\rm d.o.f.}=11/23
	\qquad\hbox{\rm and}\qquad
	\chi^2_\varphi\big/\hbox{\rm d.o.f.}=13/23.
$$
In the latter case the deviation is much more significant:
$$
	\chi^2_\psi\big/\hbox{\rm d.o.f.}=11/16
	\qquad\hbox{\rm and}\qquad
	\chi^2_\varphi\big/\hbox{\rm d.o.f.}=29/16.
$$
An approximately $60$ \%
increase in $\chi^2$ is obtained when
the fitting procedure was performed on the 11
distributions simultaneously
using variable exponent~$\mu$. Again, no significant change was observed
in the values of the fit parameters. These are
\hbox{$k=6.06\pm0.34$} and \hbox{$\mu=1.29\pm0.03$},
the exponent~$\mu$ measuring the degree of
deviation from an exponentially falling tail
proved to be smaller than in refs.~[11,12].
The scaling function $\varphi(z)$
corresponding to the fitted parameters is represented by the
solid curve in Figure~1.

Let us summarize our main results.
We have demonstrated that besides $\bar nP_n$
the more simple combination $nP_n$ also scales to a universal curve
in the variable $n/\bar n$ if KNO scaling holds valid. This somewhat
unexpected behavior follows from the fact that the second
normalization condition for $\psi(z)$, Eq.~(3), defines a second
properly normalized KNO scaling function,
\hbox{$\varphi(z)=z\,\psi(z)$}. Thus Eq.~(3) provides, on the one hand,
certain restrictions: the first moment of $\psi(z)$ is constrained
to be ${\cal M}_1=1$ and this usually generates a scale parameter
$\sigma\not=1$ for $\psi(z)$. On the other hand, Eq.~(3)
defines a KNO scaling function without the above restrictions.
The $q$th moment of $\varphi(z)$ coincides with the
$(q+1)$th moment of $\psi(z)$, further,
$\varphi(z)$ generates scale parameter $\sigma=1$
since it depends only on
the combination of $z$ and the scale parameter of $\psi(z)$.
Perhaps the most profitable feature of the new scaling function
lies in the fact that the statistical and
systematic uncertainties of $\bar n$ are not involved by the $nP_n$.
This makes $\varphi(z)$ more selective than $\psi(z)$.
According to our experience gained in the analysis of 11 multiplicity
distributions the increase in $\chi^2$ statistics if the $nP_n$
are fitted to a model prediction
instead of $\bar nP_n$ exceeds 50 \% for the majority of
distributions.
On the basis of the above features we conclude that the
analysis of the scaling
function $\varphi(z)$ may become a useful method in the
exploration of physical processes that give rise to KNO scaling
behavior for the multiplicity distributions.

\bs\bs
\noindent
{\tsc Acknowledgements\ }
I am indebted to T. Cs\"org\H o and S. Krasznovszky for the valuable
discussions. This work was supported by the Hungarian
Science Foundation under grants No. OTKA-2972 and OTKA-F4019.
\vfill\eject
\noindent
{\tsc References\ }
\ms
\frenchspacing

\item{[1]} Z. Koba, H.B. Nielsen and P. Olesen,
	Nucl. Phys. B 40 (1972) 317.

\item{[2]} A. Stuart and J.K. Ord,
	Kendall's Advanced Theory of Statistics, Vol. I,
	Distribution Theory. Charles Griffin, London 1987.

\item{[3]} P. Carruthers, A Personal Recollection of Count Statistics
	and Especially the Negative Binomial Distribution,
	Arizona Univ. Report, AZPH-TH-94-10.

\item{[4]} M. Abramowitz and I. Stegun, Handbook of Mathematical
	Functions. Dover, New York 1965.

\item{[5]} P. Carruthers and C.C. Shih,
	Int. J. Mod. Phys. A 2 (1987) 1447.

\item{[6]} T. Ferguson, Ann. Math. Statist. 33 (1962) 986.,
	see also \hfill\break
	L. Bondesson, Scand. Actuarial J. 78 (1978) 48.

\item{[7]} R. Szwed, G. Wrochna and A.K. Wr\'oblewski,
	Z. Phys. C 47 (1990) 449.
	\hfill\break
	R. Szwed, G. Wrochna and A.K. Wr\'oblewski,
	Mod. Phys. Lett. \hfill\break
	A 6 (1991) 245 and 981.

\item{[8]} N.L. Johnson and S. Kotz,
	Distributions in Statistics, Vol. II,
	Continuous Univariate Distributions.
	Houghton Mifflin, Boston 1970.

\item{[9]} A. Schenzle and H. Brand,
	Phys. Rev. A 20 (1979) 1628.

\item{[10]} S. Krasznovszky and I. Wagner,
	Nuovo Cimento A 76 (1983) 539.

\item{[11]} S. Krasznovszky and I. Wagner,
	Phys. Lett. B 295 (1992) 320.
	\hfill\break
	S. Krasznovszky and I. Wagner,
	Phys. Lett. B 306 (1993) 403.

\item{[12]} Yu. L. Dokshitzer, Phys. Lett. B 305 (1993) 295.

\item{[13]} I.M. Dremin, Quantum Chromodynamics and Multiplicity
	Distributions, \hfill\break hep-ph/9406231.

\item{[14]}
	ALEPH Collab., D. Decamp et al.,
	Phys. Lett. B 273 (1991) 181.
	\hfill\break
	AMY Collab., H.W. Zheng et al.,
	Phys. Rev. D 42 (1990) 737.
	\hfill\break
	ARGUS Collab., H. Albrecht et al.,
	Z. Phys. C 54 (1992) 13.
	\hfill\break
	DELPHI Collab., P. Abreau et al.,
	Z. Phys. C 50 (1991) 185.
	\hfill\break
	HRS Collab., M. Derrick et al.,
	Phys. Rev. D 34 (1986) 3304.
	\hfill\break
	L3 Collab., B. Adeva et al.,
	Z. Phys. C 55 (1992) 39.
	\hfill\break
	OPAL Collab., P.D. Acton et al.,
	Z. Phys. C 53 (1992) 539.
	\hfill\break
	TASSO Collab., W. Braunschweig et al.,
	Z. Phys. C 45 (1989) 193.

\bs\bs\bs\noindent
{\tsc Figure Captions\ }
\ms\noindent
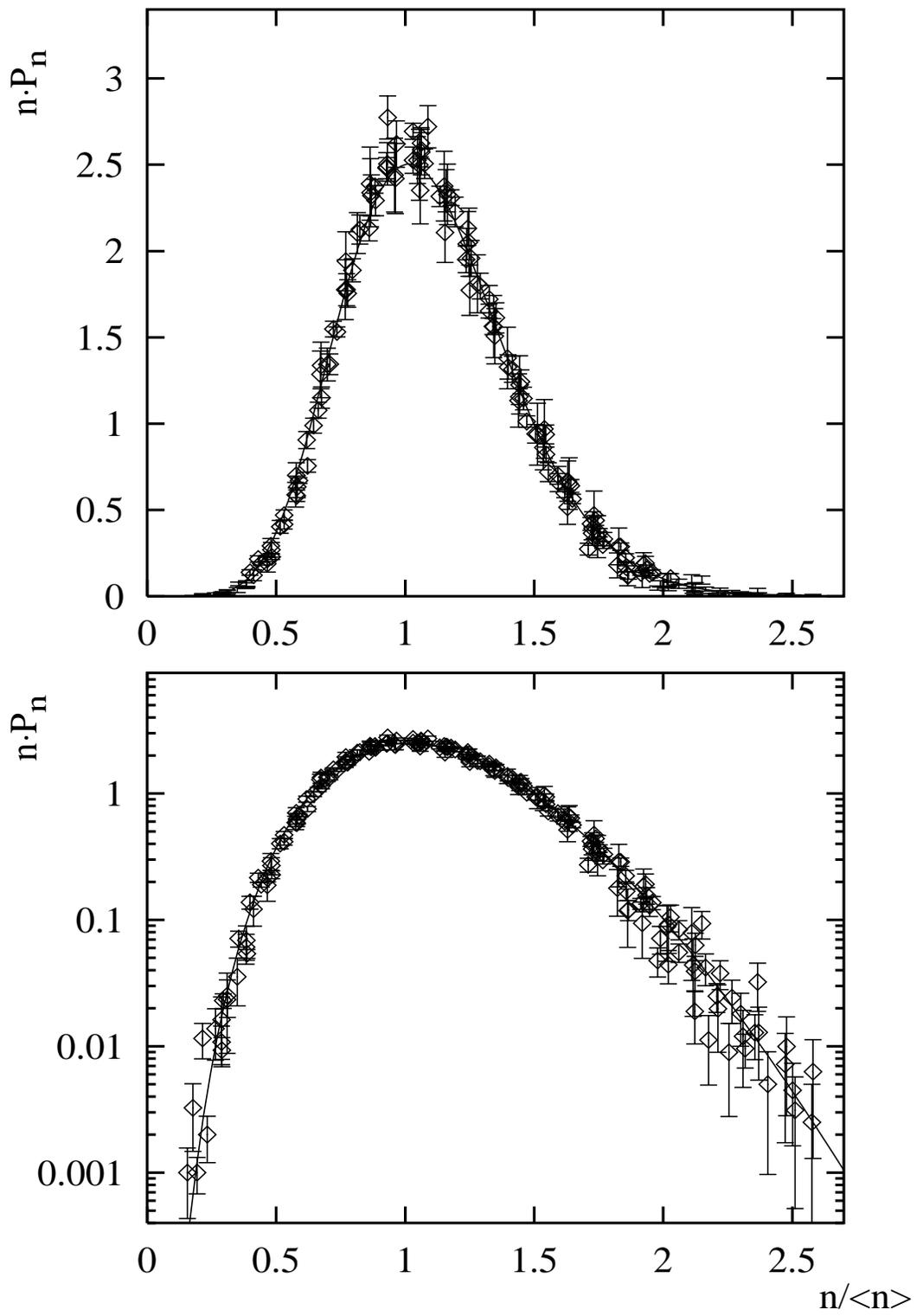
Figure 1.\ \
The KNO scaling function $nP_n(n/\bar n)=\varphi(z)$ for
charged particle multiplicity data in $e^+e^-$ annihilations.
The marked points represent ALEPH, AMY, ARGUS, DELPHI, HRS, L3,
OPAL and TASSO data in the c.m. energy range
$\sqrt s=9.4-91$ GeV.
The solid curve represents the theoretical $\varphi(z)$
corresponding to Eq.~(16), see the text for details.

\bye